\documentclass[article]{emulateapj}

\shorttitle{COSMOS Black Hole Masses}
\shortauthors{Trump et al.}

\begin{document}


\title{Observational Limits on Type 1 AGN Accretion Rate in COSMOS\altaffilmark{1}}

\author{
  Jonathan R. Trump,\altaffilmark{2}
  Chris D. Impey,\altaffilmark{2}
  Brandon C. Kelly,\altaffilmark{3}$^,$\altaffilmark{4}
  Martin Elvis,\altaffilmark{3}
  Andrea Merloni,\altaffilmark{5}
  Angela Bongiorno,\altaffilmark{5}$^,$\altaffilmark{6}
  Jared Gabor,\altaffilmark{2}
  Heng Hao,\altaffilmark{3}
  Patrick J. McCarthy,\altaffilmark{7}
  John P. Huchra,\altaffilmark{3}
  Marcella Brusa,\altaffilmark{5}
  Nico Cappelluti,\altaffilmark{5}
  Anton Koekemoer,\altaffilmark{8}
  Tohru Nagao,\altaffilmark{9}
  Mara Salvato,\altaffilmark{10}
  and Nick Z. Scoville\altaffilmark{10}
}

\altaffiltext{1}{
  Based on observations with the NASA/ESA \emph{Hubble Space
  Telescope}, obtained at the Space Telescope Science Institute, which
  is operated by AURA Inc, under NASA contract NAS 5-26555; and based
  on data collected at the Magellan Telescope, which is operated by
  the Carnegie Observatories.
\label{cosmos}}

\altaffiltext{2}{
  Steward Observatory, University of Arizona, 933 North Cherry Avenue,
  Tucson, AZ 85721
\label{Arizona}}

\altaffiltext{3}{
  Harvard-Smithsonian Center for Astrophysics, 60 Garden Street,
  Cambridge, MA 02138
\label{CfA}}

\altaffiltext{4}{
  Hubble Fellow
\label{Hubble}}

\altaffiltext{5}{
  Max Planck-Institut f\"ur Extraterrestrische Physik,
  Giessenbachstrasse 1, D-85748 Garching, Germany
\label{Max Planck}}

\altaffiltext{6}{
  University of Maryland, Baltimore County, 1000 Hilltop Circle,
  Baltimore, MD 21250
\label{Maryland}}

\altaffiltext{7}{
  Observatories of the Carnegie Institute of Washington, Santa Barbara
  Street, Pasadena, CA 91101
\label{Carnegie}}

\altaffiltext{8}{
  Space Telescope Science Institute, 3700 San Martin Drive, Baltimore,
  MD 21218
\label{STScI}}

\altaffiltext{9}{
  Ehime University, 2-5 Bunkyo-cho, Matsuyama 790-8577, Japan
\label{Ehime}}

\altaffiltext{10}{
  California Institute of Technology, MC 105-24, 1200 East California
  Boulevard, Pasadena, CA 91125
\label{Caltech}}



\def\etal{et al.}
\newcommand{\Hb}{\hbox{{\rm H}\kern 0.1em$\beta$}}
\newcommand{\MgII}{\hbox{{\rm Mg}\kern 0.1em{\sc ii}}}
\newcommand{\CIII}{\hbox{{\rm C}\kern 0.1em{\sc iii}]}}
\newcommand{\CIV}{\hbox{{\rm C}\kern 0.1em{\sc iv}}}
\newcommand{\OIII}{\hbox{[{\rm O}\kern 0.1em{\sc iii}]}}

\begin{abstract}

We present black hole masses and accretion rates for 182 Type 1 AGN in
COSMOS.  We estimate masses using the scaling relations for the broad
\Hb, \MgII, and \CIV~emission lines in the redshift ranges
$0.16<z<0.88$, $1<z<2.4$, and $2.7<z<4.9$.  We estimate the accretion
rate using an Eddington ratio $L_I/L_{Edd}$ estimated from optical and
X-ray data.  We find that very few Type 1 AGN accrete below
$L_I/L_{Edd} \sim 0.01$, despite simulations of synthetic spectra
which show that the survey is sensitive to such Type 1 AGN.  At lower
accretion rates the BLR may become obscured, diluted or nonexistent.
We find evidence that Type 1 AGN at higher accretion rates have higher
optical luminosities, as more of their emission comes from the cool
(optical) accretion disk with respect to shorter wavelengths.  We
measure a larger range in accretion rate than previous works,
suggesting that COSMOS is more efficient at finding low accretion rate
Type 1 AGN.  However the measured range in accretion rate is still
comparable to the intrinsic scatter from the scaling relations,
suggesting that Type 1 AGN accrete at a narrow range of Eddington
ratio, with $L_I/L_{Edd} \sim 0.1$.

\end{abstract}


\section{Introduction}

Supermassive black holes (SMBHs) reside in almost all local galaxies
\citep{kor95, ric98}.  The mass of the SMBH is observed to be tightly
correlated with the mass, luminosity, and velocity dispersion of the
host galaxy bulge \citep[e.g.,][]{mag98, geb00, fer00}.  SMBHs grow by
accretion as active galactic nuclei (AGN), and all massive galaxies
have one or more of these active phases \citep{sol82, mag98, mar04}.
More luminous AGN are observed to peak at higher redshift
\citep{ued03, bra05, bon07}, exhibiting ``downsizing'' by analogy to
the preference of luminous and massive galaxies to form at high
redshifts.  Both downsizing and the correlations between SMBH and the
host bulge suggest that the growth of AGN and the formation of
galaxies are directly connected through feedback \citep{sil98, dim05}.

Understanding the role of AGN in galaxy evolution requires
measurements of SMBH mass and accretion over the cosmic time.  The
SMBH mass can be directly estimated by modeling the dynamics of nearby
gas or stars, but this requires high spatial resolution and is limited
to HST observations of nearby galaxies.  Reverberation mapping of Type
1 AGN (with broad emission lines) uses the time lag between
variability in the continuum and the broad line region (BLR) to
estimate the radius of the broad line region, $R_{BLR}=ct_{lag}$
\citep[for a review, see][]{pet06}.  Then, if the broad line region
virially orbits the source of the continuum emission, the SMBH mass is
$M_{BH}=fR_{BLR}v_{fwhm}^2$, where $f$ represents the unknown BLR
geometry and $v_{fwhm}$ is the velocity width of the broad emission
line.  This technique has many potential systematic errors
\citep{kro01, mar08}, but its mass estimates agree with those from
dynamical estimators \citep{dav06, onk07} and those from the
$M_{BH}$-$\sigma*$ correlation \citep{onk04, gre06}.  In principle,
reverberation mapping can be applied to AGN at any redshift, but in
practice, the need for many periodic observations has limited
reverberation mapping mass estimates to only $\sim$35 local AGN.

Instead, the vast majority of AGN mass estimates have come from a set
of scaling relations.  Reverberation mapping data led to the discovery
that $R_{BLR}$ correlates with the continuum luminosity \citep{kas00},
with $R_{BLR} \sim L^{\alpha}$, where $\alpha \sim 0.5$ \citep{ben06,
kas07}.  This allows for estimates of $M_{BH}$ from single epoch
spectra with scaling relations:
\begin{equation}
  \log \left( \frac{M_{BH}}{M_{\odot}} \right) = 
  A + B \log(\lambda L_{\lambda}) + 2 \log(v_{FWHM})
\end{equation}

Some authors replace the continuum luminosity with the recombination
line luminosity \citep{wu04} or the FWHM with the second moment
$\sigma$ \citep{col06}, yielding minor systematic differences in
estimated $M_{BH}$ \citep{shen08}.  While these scaling relations are
based upon reverberation mapping of only local AGN, the method is
based upon the ability of the central engine to ionize the broad line
region \citep{kas00}, and there is no physical reason to suggest that
the ionization of AGN should evolve with redshift \citep{die04,
ves04}.  Thus the scaling relations can be used to study the
distribution and evolution of Type 1 AGN masses.  \citet{kol06} showed
that Type 1 AGN tend to accrete at a narrow range of Eddington ratio,
typically $0.01L_{Edd}<L<L_{Edd}$.  \citet{kol06} suggest a minimum
accretion rate for Type 1 AGN, with AGN of lower accretion rate
observed as ``naked'' Type 2 AGN without a broad line region
\citep{hop08}.  \citet{gav08} additionally suggest that lower
luminosity Type 1 AGN accrete less efficiently than brighter quasars.

In this work we report black hole masses and study the demographics of
182 Type 1 AGN in the Cosmic Evolution Survey
\citep[COSMOS,][]{sco07}.  We introduce the data and outline our
spectral fitting in \S 2, and discuss the black hole masses, their
associated errors, and our completeness in \S 3.  We close with
discussion of Type 1 AGN accretion rates in \S 4.  All luminosities
are calculated using $h=0.70$, $\Omega_M=0.3$, $\Omega_{\Lambda}=0.7$.

\section{Observational Data}

\subsection{Sample}

The Cosmic Evolution Survey \citep[COSMOS, ][]{sco07} is a 2 deg$^2$
HST/ACS survey \citep{koe07} with ancillary deep multiwavelength
observations.  The depth of COSMOS over such a large area is
particularly suited to the study of low-density, rare targets like
active galactic nuclei (AGN).  The most efficient way to select AGN is
by their X-ray emission, and XMM-{\it Newton} observations of COSMOS
\citep{cap08} reach fluxes of $1 \times 10^{-15}$ cgs and $6 \times
10^{-15}$ cgs in the 0.5-2 keV and 2-10 keV bands, respectively.  The
matching of X-ray point sources to optical counterparts is described
by \citet{bru07} and \citet{bru09}.  \citet{tru09} performed a
spectroscopic survey of XMM-selected AGN in COSMOS, revealing 288 Type
1 AGN with X-ray emission and broad emission lines in their spectra.
Here we investigate a chief physical property, the black hole mass,
for the Type 1 AGN in this survey.

From the \citet{tru09} sample, we choose the 182 Type 1 AGN with
high-confidence redshifts and with \Hb, \MgII, or \CIV~present in the
observed wavelength range.  That is, we select only $z_{\rm conf} \ge
3$ AGN, empirically determined by \citet{tru09} to be at least 90\%
likely to have the correct classification and redshift.  All of the
Type 1 AGN spectra are dominated by blue power-law continua, with no
obvious (beyond the noise) absorption line signature from the host
galaxy.  The broad emission line requirement restricts us to the
redshift ranges $0.16<z<0.88$, $1<z<2.4$, and $2.7<z<4.9$.  At these
redshifts the AGN spectroscopy is $>90\%$ complete to $i_{\rm
AB}^+<22$ \citep[see Figure 13 of][]{tru09}, where $i_{\rm AB}^+$ is
the AB magnitude from the COSMOS CFHT observations.  Our ability to
measure a broad line width, however, is a slightly more complicated
function of the spectral signal to noise.  We characterize our
completeness as a function of broad line width and S/N in \S 3.2.

The majority (133) of the 182 AGN have spectra from Magellan/IMACS
\citep{big98}, with wavelength coverage from 5600-9200\AA~and
$\sim$10$\AA$~resolution.  The remaining bright 49 quasars have
publicly available spectra from the Sloan Digital Sky Survey (SDSS)
quasar catalog \citep{schn07}, with 3800-9200\AA~wavelength coverage
and $\sim$3\AA~resolution.  The resolution of both surveys is more
than sufficient for measuring broad emission line widths; our
narrowest broad emission line is 1300 km/s wide, compared to the
resolution limits of IMACS ($\sim$600 km/s) and the SDSS ($\sim$200
km/s).

\subsection{Spectral Fitting}

The optical/UV spectrum of a Type 1 AGN can be roughly characterized
as a power-law continuum, $f_{\nu} \propto \nu^{-\alpha}$, with
additional widespread broad iron emission and broad emission lines
\citep{van01}.  We followed \citet{kel08} to model the spectra, using
an optical Fe template from \citet{ver04} and a UV Fe template from
\citet{ves01}.  Each spectrum was simultaneously fit with a power-law
continuum and an iron template using the Levenberg-Marquardt method
for nonlinear $\chi^2$ minimization.  We calculated the continuum
luminosity from the power-law fit parameters.  Light from the host
galaxy can artificially inflate the estimated AGN continuum
luminosity, as shown by \citet{ben08} for the 35 AGN with
reverberation mapping data.  In particular, \citet{ben08} find that at
5100\AA, host galaxies contribute $\sim$20\% of the measured flux for
AGN with ${\lambda}L_{5100}>10^{44}$ erg/s, and $\sim$45\% of the flux
for AGN of ${\lambda}L_{5100}<10^{44}$ erg/s.  We note that the
expected host contamination at $L_{3000}$ or $L_{1350}$ is much
smaller, since most host galaxies (excepting very active star-forming
hosts) have much less flux blueward of the 4000\AA~break.  So for the
majority of our sample, the 150 AGN with masses measured from \MgII~or
\CIV, we do not expect host contamination to be significant.  However,
the 15 AGN with ${\lambda}L_{3000}<10^{44}$ erg/s and \Hb-derived
masses may have luminosity estimates overestimated by a factor of two,
leading to black hole masses and Eddington ratios systematically
overestimated by $\sim$0.15 dex.  Future work in COSMOS will use host
decompositions from HST/ACS images of $z<1$ AGN to better characterize
host contamination, but in this work we make no corrections for host
galaxy flux.

To fit the broad emission lines we subtracted the continuum and Fe
emission fits.  Narrow absorption lines and narrow emission lines near
the broad line (e.g., \OIII$\lambda$4959 and \OIII$\lambda$5007 near
\Hb) were fit by the sum of 1-2 Gaussian functions and removed.  Again
following \citet{kel08}, each remaining broad emission line profile
was fit by the sum of 1-3 Gaussian functions, minimizing the Bayesian
information criterion \citep[BIC,][]{schw79}.  Roughly, 2 Gaussians
provided the best fit for $\sim$40\% of line profiles, while 1 or 3
Gaussians each provided the best fit for $\sim$30\% of line profiles.
All fitting was interactive and inspected visually, and if the
multiple-Gaussian fit revealed a narrow ($<600$ km/s) line in an
\Hb~emission line, the component was attributed to non-BLR origins and
was removed.  The FWHM was calculated directly from the
multiple-Gaussian fit in order to minimize the effects of noise in the
original spectra.  The multiple-Gaussian fit was robust to a variety
of line profiles, and simulated spectra (see \S3.2) revealed measured
FWHM errors of only $\sigma_{FWHM}/FWHM \sim 10\%$.

\begin{figure}
\scalebox{1.2}
{\plotone{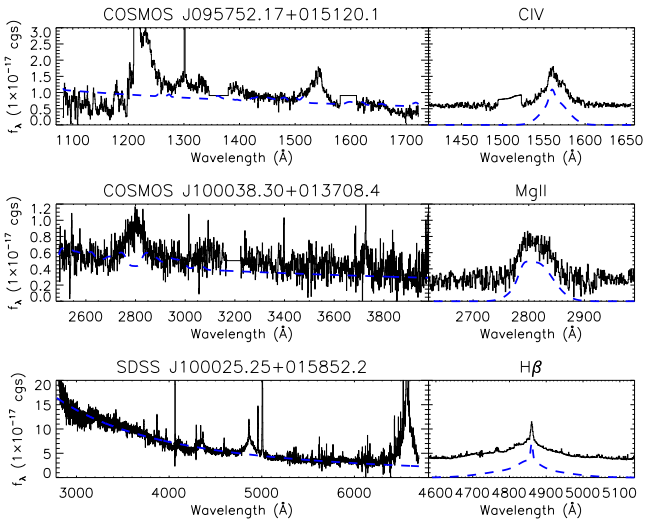}}
\figcaption{Three spectra, representing fits and line profiles for
Type 1 AGN with \CIV~(top), \MgII~(middle), and \Hb~(bottom).  In the
left panels, the spectra are shown by black histograms, and the dashed
blue lines are the power-law plus iron emission continuum fits.  At
right, the blue dashed line shows the multiple-Gaussian fit to the
continuum-subtracted line profiles, which are shown in black and
shifted above for clarity.  The fit to the \Hb~line profile in the
lower right panel includes a narrow ($\sigma=433$ km/s) Gaussian which
was removed before calculating the broad emission line width.  The top
two spectra were observed with Magellan/IMACS and have
$\sim$10$\AA$~resolution, while the bottom spectrum comes from the
SDSS and has $\sim$3$\AA$~resolution.
\label{fig:fits}}
\end{figure}

Three examples of spectra with fitted continua and
continuum-subtracted line profiles are shown in Figure \ref{fig:fits}.
The spectra are representative of typical fits for each of the \CIV,
\MgII, and \Hb~emission lines.  At left the power-law and iron
emission fits are shown by the dashed blue lines.  The right panel
shows the multiple-Gaussian line profile fits as dashed blue lines,
with the continuum-subtracted line profile shifted above by an
arbitrary amount for clarity.  The fit to the \Hb~line profile in the
bottom right panel includes a narrow ($\sigma=433$ km/s) Gaussian
which is not associated with the BLR and was removed.  Even for noisy
spectra like the middle panel, the spectral fitting provides a robust
continuum and isolates the emission line.

\section{Estimated Black Hole Masses}

We estimate black hole masses using our measured broad line velocity
widths and the scaling relations of \citet{ves09} for \MgII~and
\citet{ves06} for \Hb~and \CIV.  These relations all take the form of
Equation 1, with ${\lambda}L_{\lambda}$ in units of $10^{44}$~erg/s
and $v_{FWHM}$ in units of 1000~km/s; $A=6.91$, $B=0.50$, and
$\lambda=5100$\AA~for \Hb; $A=6.86$, $B=0.50$, and
$\lambda=3000$\AA~for \MgII; $A=6.66$, $B=0.53$, and
$\lambda=1350$\AA~for \CIV.  The \MgII~relation was derived from SDSS
quasars with both \CIV~and \MgII~in the spectrum, and it is designed
to produce black hole masses consistent with those measured from \CIV.
In our sample, we measure \Hb~for 32 AGN, \MgII~for 134 AGN, and
\CIV~for 38 AGN (19 SDSS AGN have both \MgII~and \CIV, and 3 have both
\Hb~and \MgII).  AGN with estimates of $M_{BH}$ from two different
emission lines are treated as two separate objects in our subsequent
analyses.

\begin{figure}
\scalebox{1.2}
{\plotone{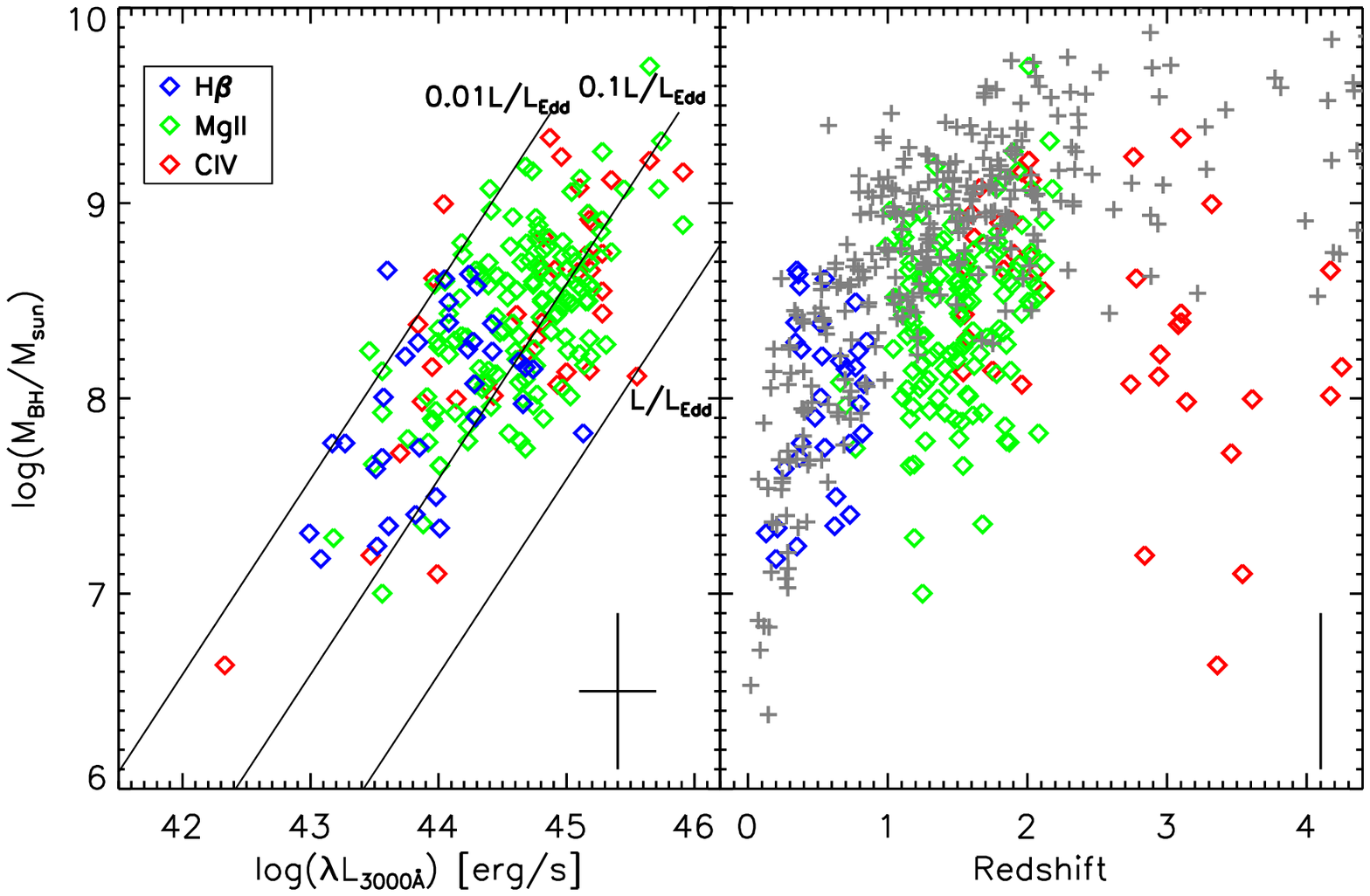}}
\figcaption{Black hole mass versus optical luminosity and redshift for
182 Type 1 AGN in COSMOS.  The black hole masses are derived from the
broad line velocity widths of \Hb~(blue diamonds), \MgII~(green
diamonds), and \CIV~(red diamonds) depending on the redshift
\citep{ves06, ves09}.  The black hole masses have uncertainties of
$\sim$0.4 dex from the scaling relations \citep{kro01, shen08}, as
shown by the typical error bars are shown in the lower right of each
panel (the redshift error is $\lesssim 1\%$).  The diagonal tracks at
left represent Eddington ratios, assuming a bolometric correction of
$L_{bol}=5({\lambda}L_{3000})$ \citep{ric06}.  At right the gray
crosses show scaling relation masses of SDSS quasars \citep{kel08}.
The depth of COSMOS allows us to probe low mass and weakly accreting
SMBHs, revealing that $L/L_{Edd} \lesssim 0.01$ Type 1 AGN do not
exist.
\label{fig:plotmbh}}
\end{figure}

Table \ref{tbl:mbh} presents the catalog of black hole masses and line
measurements.  AGN with both \MgII~and \CIV~or both \Hb~and
\MgII~present have two entries in Table \ref{tbl:mbh}, one for each
emission line.  The full catalog then contains 204 entries: 182 Type 1
AGN with $M_{BH}$ estimates, 22 of which have two sets of broad
emission line measurements.  The black hole masses are shown with
continuum luminosity (calculated from the power-law fit) and redshift
in Figure \ref{fig:plotmbh}.  The diagonal tracks in the figure
represent Eddington ratios using a bolometric correction of 5 for
${\lambda}L_{3000}$ \citep{ric06}.  We also show a comparison
sample of brighter SDSS quasars \citep{kel08} in order to highlight
the lower black hole masses probed by COSMOS.

\begin{deluxetable*}{lcrcccrc}
\tablecolumns{8}
\tablewidth{0pc}
\tablecaption{COSMOS Type 1 AGN Black Hole Mass Catalog
\label{tbl:mbh}}
\tablehead{
  \colhead{Object} &
  \colhead{Redshift} &
  \colhead{S/N} &
  \colhead{$\log({\lambda}L_{3000\AA})$} &
  \colhead{$\log(L_{0.5-2 \rm keV})$} &
  \colhead{Line} &
  \colhead{FWHM} &
  \colhead{$\log(M_{BH})$} \\
  \colhead{(J2000)} &
  \colhead{} &
  \colhead{(per pixel)\tablenotemark{a}} &
  \colhead{[erg/s]} &
  \colhead{[erg/s]\tablenotemark{b}} &
  \colhead{} &
  \colhead{(km/s)} &
  \colhead{$[M_{\odot}]$} }
\startdata
  SDSS J095728.34+022542.2 & 1.54 &  7.0 & 45.00 & 44.47 & MgII & 4491 & 8.665 \\
  SDSS J095728.34+022542.2 & 1.54 &  7.0 & 45.00 & 44.47 & CIV &  2776 & 8.135 \\
COSMOS J095740.78+020207.9 & 1.48 & 17.9 & 43.46 & 44.41 & MgII & 6701 & 8.244 \\
  SDSS J095743.33+024823.8 & 1.36 &  3.4 & 44.60 & 43.50 & MgII & 3472 & 8.243 \\
COSMOS J095752.17+015120.1 & 4.17 &  7.3 & 45.19 & 44.36 & CIV &  4603 & 8.656 \\
COSMOS J095753.49+024736.1 & 3.61 &  4.8 & 44.14 & 44.53 & CIV &  2629 & 7.997 \\
  SDSS J095754.11+025508.4 & 1.57 &  6.1 & 45.07 & 44.45 & MgII & 4500 & 8.701 \\
  SDSS J095754.70+023832.9 & 1.60 &  8.0 & 45.17 & 43.69 & MgII & 3361 & 8.498 \\
  SDSS J095754.70+023832.9 & 1.60 &  8.0 & 45.17 & 43.69 & CIV &  6384 & 8.946 \\
  SDSS J095755.08+024806.6 & 1.11 &  8.7 & 44.92 & 44.08 & MgII & 3574 & 8.426 \\
COSMOS J095755.34+022510.9 & 2.74 &  3.2 & 44.38 & -1.00 & CIV &  3879 & 8.074 \\
COSMOS J095755.48+022401.1 & 3.10 & 19.9 & 45.28 & 45.00 & CIV &  3527 & 8.436 \\
\enddata
\tablenotetext{a}{The SDSS spectra have 3 pixels per resolution
element, and the Magellan/IMACS spectra have 5 pixels per resolution
element.}
\tablenotetext{b}{AGN with no soft X-ray detection have an entry of
-1.00 for $\log(L_{0.5-2 \rm keV})$.}
\end{deluxetable*}

\subsection{Error}

\begin{figure}
\scalebox{1.2}
{\plotone{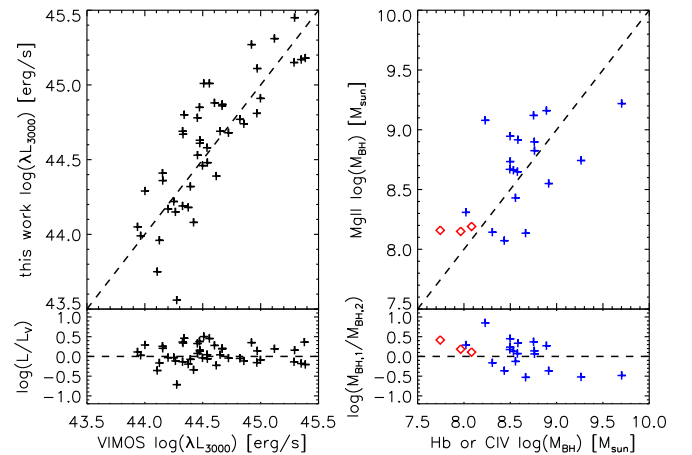}}
\figcaption{At left, the luminosity estimates for 48 AGN in this work
are compared to estimates of the same objects from \citet{mer09},
which have independent redshifts from VLT/VIMOS spectra and are
calculated from a fit to the multiwavelength SED.  The scatter between
the two luminosity estimates is $\sigma=0.25$ dex, which contributes
very little to the overall scatter in $M_{BH}$.  At right, $M_{BH}$
estimates are compared for the 22 spectra with two emission lines
(either \MgII~and \Hb, red diamonds, or \MgII~and \CIV, blue crosses).
The scatter between separate $M_{BH}$ estimates is only $\sigma=0.36$
dex, indicative that the intrinsic scatter of $\sim$0.4 dex dominates
the $M_{BH}$ error.
\label{fig:comparecheck}}
\end{figure}

The scaling relations have uncertainties of $\sim$0.4 dex, although
there may be larger systematic uncertainties \citep{kro01, col06,
fin08, mar08}.  Measurement errors in the luminosity and emission line
FWHM also contribute, but the uncertainty from the scaling relations
dominates.  We test the luminosity error in the left panel of Figure
\ref{fig:comparecheck}, which compares the luminosity estimates from
this work to duplicate estimates from \citet{mer09}.  The luminosity
estimates of \citet{mer09} use independent redshifts from VLT/VIMOS
spectra and are calculated from a fit to the IR to X-ray
multiwavelength spectral energy distribution, instead of from the
optical spectrum itself (as in this work).  The scatter between the
two luminosity estimates is $\sigma=0.25$ dex.  The Type 1 AGN in
COSMOS have an average variability of $\sim$0.15 dex \citep{sal09},
the remaining luminosity scatter can be attributed to the different
methods of estimates.  Since $M_{BH} \sim L^{0.5}$, our luminosity
error contributes very little to the overall $M_{BH}$ uncertainty.

Line measurements of synthetic spectra (described in \S 3.2 below)
show that our FWHM error is only $\sigma_{FWHM}/FWHM \sim 10\%$ at
$i_{\rm AB}^+ \sim 22$.  The right panel of Figure
\ref{fig:comparecheck} compares the duplicate estimates of $M_{BH}$
for spectra with two broad emission lines.  Red diamonds indicate
spectra with both \MgII~and \Hb, while blue crosses indicate both
\MgII~and \CIV.  The scatter between the different estimates is only
$\sigma=0.36$ dex, nearly the same as the expected intrinsic scatter
for $M_{BH}$.  This suggests that the statistical error in the mass
estimators is not correlated to the choice of emission line \citep[see
also][]{kel07}.  If there were systematic offsets in the mass
estimators, they would cause a constant shift in the mass estimate for
each line, and therefore would not contribute to the scatter between
two lines.  The statistical intrinsic scatter, however, would not
``cancel'' in such a way.  Because the scatter between lines is
comparable to the expected intrinsic scatter, our estimates of
$M_{BH}$ probably do not have significant systematic errors.

\subsection{Completeness}

Previous work \citep{tru09} tested the completeness of the Type 1 AGN
sample, with simulated spectra showing that Type 1 AGN are correctly
identified (with high confidence redshifts) at 90\% completeness to
S/N$\gtrsim$2.87.  But even for correctly identified Type 1 AGN, our
measurements of $M_{BH}$ are roughly limited by spectral S/N and FWHM,
since we cannot identify or measure broad emission lines for spectra
that have lines so broad that they become confused with noise or the
Fe emission.  To test these limits, we create 100 synthetic spectra of
Type 1 AGN with \MgII~in the observed wavelength range.  We choose
\MgII~because it is the most common line used for our $M_{BH}$
calculations and also because it is the broad emission line most
contaminated by widespread iron emission.  These synthetic spectra are
formed by making a composite of all $1<z<2.4$ observed spectra, then
removing the rest-frame $2700\AA<\lambda<2900\AA$ \MgII~region.  To
each of the 100 spectra we then re-add a \MgII~region with random
FWHMs and line areas, and then each spectrum has random noise added.
We choose the line areas and noise to be normally distributed in the
ranges of measured line area and S/N in the original spectra, while
the FWHMs are chosen to probe our sensitivity to the broadest line
widths, FWHM$\gtrsim 10000$ km/s.

\begin{figure}
\scalebox{1.2}
{\plotone{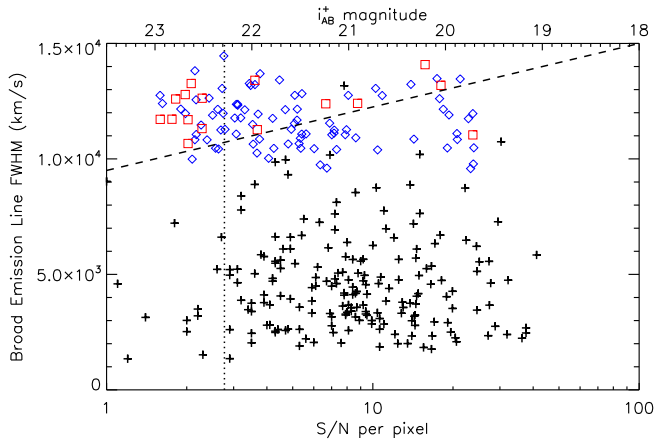}}
\figcaption{The broad emission line FWHM and S/N for our 182 Type 1
AGN are shown as black crosses.  The upper axis shows the $i_{\rm
AB}^+$ magnitude scale which roughly corresponds to the S/N
\citep[from Figure 11 of][]{tru09}.  Also shown are the measured FWHM
and S/N for 100 simulated Type 1 AGN spectra: blue diamonds represent
spectra we can correctly identify and measure, while red squares
represent synthetic spectra that we incorrectly identify or cannot
properly measure FWHM.  The dashed line shows the limit to which we
correctly identify and measure $>95\%$ of synthetic spectra.  The
dotted vertical line shows the 90\% completeness to identifying and
assigning high-confidence redshifts to Type 1 AGN, from additional
simulations \citep{tru09}.  The lack of high-FWHM AGN identified in
COSMOS is not from selection effects, since we can correctly identify
and measure synthetic Type 1 AGN with much broader lines at S/N$>3$.
\label{fig:fwhmsn}}
\end{figure}

We show the FWHM and S/N of both the observed and simulated spectra in
Figure \ref{fig:fwhmsn}.  The blue diamonds and red squares represent
the simulated spectra that we successfully measure and those we miss,
respectively.  The vertical dotted line shows the sample's 90\%
completeness limit for correctly identifying high-confidence Type 1
AGN.  The dashed line shows the limit to where we can measure broad
emission lines, corresponding to $>95\%$ completeness since only one
``missed'' synthetic spectrum lies below the line.  The observed Type
1 AGN tail off well before the dashed line.  Translating FWHM and
magnitude into black hole mass, as an example, a Type 1 AGN with
$i_{\rm AB}^+ \sim 22$ at $z \sim 1.7$ with a \MgII~profile of
FWHM$=10000$ km/s would have $\log(M_{BH}/M_{\odot}) \sim 10.6$ and
$L/L_{Edd} \sim 0.0002$.  We do not detect such objects in our sample,
yet our simulations show that they do not lie beyond our detection
limits.

\section{Discussion}

In Figure \ref{fig:plotmbh} all Type 1 AGN lie within the region of
$0.01 \lesssim L/L_{Edd} \lesssim 1$, a result supported by
\citet{kol06}.  This implies that the broad emission line region of
Type 1 AGN might become undetectable as the accretion drops below $L
\sim 0.01L_{Edd}$.  Such objects might be observed as unobscured Type
2 AGN, the possible remnants of ``dead'' Type 1 AGN whose accretion
disk geometries changed as their accretion rates fell
\citep[e.g.,][]{hop08}.  Or these low accretion rate AGN may be
diluted, with their emission falling below the light of their host
galaxy.  The AGN emission may also be unable to blow out local
obscuring material, causing their BLR to lie undetected behind
obscuration.

\begin{figure*}[t]
\begin{center}
\plotone{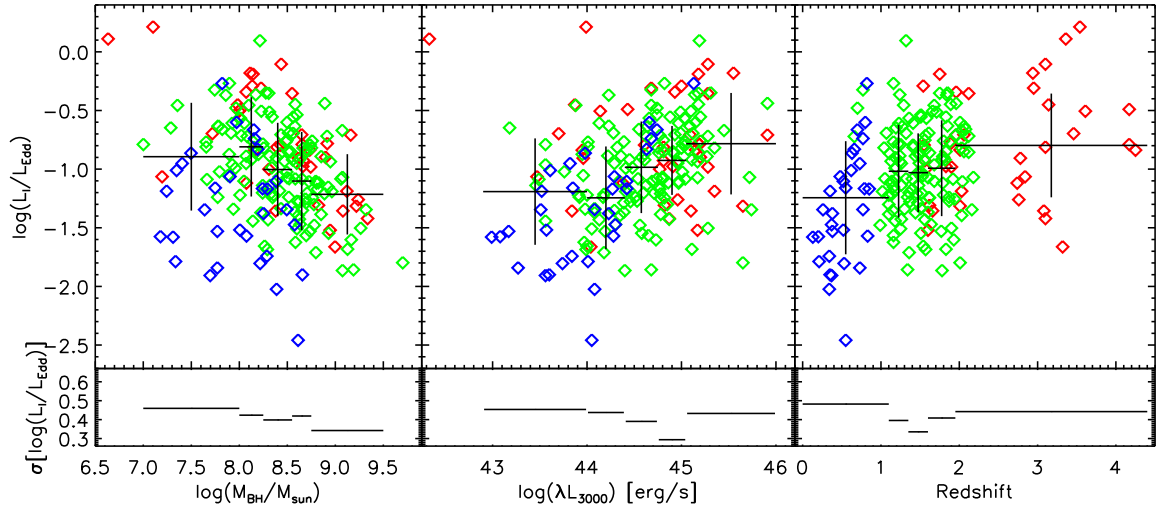}
\figcaption{The Eddington ratio (accretion rate) with black hole mass,
luminosity, and redshift for our Type 1 AGN.  Eddington ratio was
calculated using an intrinsic luminosity $L_I$ estimated from
${\lambda}L_{3000}$ and $L_{\rm 0.5-2keV}$ using the relations of
\citet{mar04}.  Diamonds represent individual objects with masses
estimated from \Hb~(blue), \MgII~(green), and \CIV~(red).  The large
crosses in the top plots show the mean accretion rate in each bin of
$M_{BH}$ or redshift, while the gray lines at the bottom show the
standard deviation (the square root of the second moment) in each bin.
The dispersion deviation is also shown by the vertical error bar in
the top plots.  Bins were chosen to each have the same number of
objects.  Selection effects cause the apparent trends of $L_I/L_{Edd}$
decreasing with $M_{BH}$ and increasing with redshift, but the
increase of $L_I/L_{Edd}$ with ${\lambda}L_{3000}$ is a physical
effect caused by changes in the accretion disk.  The dispersion is
generally $\sim$0.4 dex, higher than that of previous, less sensitive
surveys.
\label{fig:ledd}}
\end{center}
\end{figure*}

To study the accretion rates of our AGN, we calculate the intrinsic
luminosity from our measured ${\lambda}L_{3000}$ and $L_{\rm
0.5-2keV}$, using the relations of \citet{mar04}:
\begin{equation}
  \log[L_I/({\lambda}L_{3000})] = 0.65 - 0.067\mathcal{L} + 0.017\mathcal{L}^2 - 0.0023\mathcal{L}^3
\end{equation}
\begin{equation}
  \log(L_I/L_{0.5-2}) = 1.65 + 0.22\mathcal{L} + 0.012\mathcal{L}^2 - 0.0015\mathcal{L}^3
\end{equation}
Here $\mathcal{L}=\log(L_I)-45.58$ and all luminosities are in units
of erg/s.  The intrinsic luminosity $L_I$ is designed to be a
bolometric luminosity which excludes reprocessed (IR) emission, so
that the Eddington ratio $L_I/L_{Edd}$ represents a robust measure of
the accretion onto the black hole.  We use the Newton method to solve
each equation for $L_I$ from ${\lambda}L_{3000}$ and $L_{0.5-2}$.  We
then average the two values of $L_I$ for our final value (excepting 7
AGN where we estimate $L_I$ from ${\lambda}L_{3000}$ only because they
lack soft X-ray detections).

We show the Eddington ratios of our Type 1 AGN with $M_{BH}$,
${\lambda}L_{3000}$, and redshift in Figure \ref{fig:ledd}.  The
diamonds show individual objects, while the solid lines show the means
and scatter in equal-sized bins.  The scatter (the standard deviation
of the mean) is calculated as the square root of the second moment of
the data in each bin.  Our mean Eddington ratio for all Type 1 AGN is
$L_I/L_{Edd} \sim 0.1$, lower than the value of $L_I/L_{Edd} \sim 0.3$
found in previous surveys \citep{kol06,gav08}.  This is partly
explained by the depth of COSMOS: \citet{kol06} noted that their $R
\le 21.5$ AGES sample was only complete to $L_I/L_{Edd} \sim 0.1$,
while our simulations in \S 3.2 show that COSMOS can reach much weaker
accretors.  In addition, the center panel of Figure \ref{fig:ledd}
shows that accretion rate may increase with optical/UV luminosity,
suggesting an additional reason for our lower mean Eddington ratio:
most of the \citet{kol06} and \citet{gav08} AGN have
${\lambda}L_{3000}>10^{45}$ erg/s, where our AGN have $L_I/L_{Edd}
\sim 0.2$.  The majority of our AGN have ${\lambda}L_{3000}<10^{45}$
erg/s and so we find a lower mean accretion rate.

The apparent decrease in accretion rate with black hole mass and the
apparent increase in accretion rate with redshift can be explained by
selection effects: low accretion rate AGN are more difficult to detect
if they are also low mass or at higher redshift.  The increase in
accretion rate with optical luminosity, however, is also observed by
\citet{gav08} and is probably a physical effect.  We performed a
linear regression analysis of the correlation between accretion rate
and optical luminosity using the publicly available IDL program {\tt
linmix\_err.pro} \citep{kel07a}.  Using errors of 0.25 dex in
$\log({\lambda}L_{3000})$ and 0.4 dex in $\log(L_I/L_{Edd})$, linear
regression indicates that $\log(L_I/L_{Edd}) \sim (0.28 \pm
0.06)\log({\lambda}L_{3000})$.  In other words, accretion rate is
correlated with opticaly luminosity at the 4.8$\sigma$ level.
As a Type 1 AGN increases in accretion rate, its optical emission
becomes a larger fraction of its total bolometric output because its
cool accretion disk emits more brightly.  This is consistent with the
results of \citet{kel08}, which show that $\alpha_{OX}$ (the ratio
between optical/UV and X-ray flux) becomes more X-ray quiet with
accretion rate.  Thus a more rapidly accreting Type 1 AGN has more of
its emission in its cool (optical) disk than in its hot (X-ray)
corona, possibly because the disk grows larger or thicker as the
accretion rate approaches the Eddington limit.

The scatter (square root of the second moment) in $L_I/L_{Edd}$, shown
in the bottom panels of Figure \ref{fig:ledd}, is typically only
$\sim$0.4 dex in each bin.  This is greater than previously measured
dispersions \citep{kol06,gav08,fin08}, and indicates that COSMOS is
more sensitive to low accretion rate Type 1 AGN than previous studies.
Yet it is remarkable that the dispersion is not larger than the
scatter from the scaling relations: the intrinsic dispersion in
Eddington ratio might then be $\sim$0, with nearly all Type 1 AGN of a
given mass, luminosity, and/or redshift accreting at a very narrow
range of of accretion rates.  \citet{fin08} note that at such low
measured dispersions, if the intrinsic dispersion in accretion rate is
much greater than 0, then the BLR cannot be in a simple virial orbit.
Accurate scaling relations would then require a luminosity-dependent
ionization parameter \citep{mar08} or a more complex BLR geometry
\citep{fin08}.  We note, however, that the myriad uncertainties
involved in estimating $L_I/L_{Edd}$ make concrete conlusions
difficult.  And although we find significant evidence that
$L_I/L_{Edd}<0.01$ Type 1 AGN do not exist (or are very rare), the
spectroscopic flux limit may still miss some $L_I/L_{Edd}\sim0.01$ AGN
at lower luminosities.

\section{Summary}

The black hole masses of Type 1 AGN in COSMOS indicate that Type 1 AGN
accrete at a narrow range of high efficiencies, $L_I/L_{Edd} \sim
0.1$.  When the accretion rate of an AGN lowers, less of its
luminosity is emitted optically.  When a Type 1 AGN accretion rate
drops below $L_I/L_{Edd} \sim 0.01$ the BLR becomes invisible, due to
obscuration, dilution, or an altered accretion disk geometry.  We
additionally measure higher dispersions in accretion rate than
previous, less sensitive surveys, although the dispersion is still no
larger than the intrinsic uncertainty in the scaling relations.
\citet{kel08} find that the bolometric correction depends on black
hole mass, and \citet{vas09} find that it correlates with Eddington
ratio.  This makes characterizing the distributions of $L_I/L_{Edd}$
and its scatter rather difficult.  We partially mitigate the
systematic uncertainties by using both ${\lambda}L_{3000}$ and
$L_{0.5-2}$ to estimate $L_I$.  Future work in COSMOS will use more
accurate bolometric luminosities calculated from the full
multiwavelength dataset.

\acknowledgements

We thank Brad Peterson for helpful discussions on emission line
measurements and Misty Bentz for comments on host contamination.  We
thank Marianne Vestergaard for providing the Fe templates and for
discussions on the scaling relations.  Amy Stutz and Aleks
Diamond-Stanic provided useful discussions on statistics.  JRT
acknowledges support from NSF ADP grant NNX08AJ28G and an ARCS
fellowship.  BCK acknowledges support from NASA through Hubble
Fellowship grant \#HF-01220.01 awarded by the Space Telescope Science
Institute, which is operated by the Association of Universities for
Research in Astronomy, Inc., for NASA, under contract NAS 5-26555.

\begin{center}


\end{center}

\end{document}